# Creation, Chaos, Time :
# from Myth to Modern Cosmology

Jean-Pierre Luminet

Laboratoire d'Astrophysique de Marseille, Aix-Marseille Université,
CNRS-UMR 7326, 13388 Marseille, France
& Laboratoire Univers et Théories, Observatoire de Paris, 92195 Meudon,
France

## Abstract

Every society has a story rooted in its most ancient traditions, of how the earth and sky originated. Most of these stories attribute the origin of all things to a Creator - whether God, Element or Idea. We first recall that in the Western world all discussions of the origin of the world were dominated until the 18th century by the story of *Genesis,* which describes the Creation as an ordered process that took seven days. Then we show how the development of mechanistic theories in the 18th century meant that the idea of an organized Creation gave way to the concept of evolution, helped by the fact that in the 19th century astrophysicists discovered that stars had their origin in clouds of gas. We conclude with Big bang theory, conceived at the beginning of the 20th century, that was subsequently developed into a more or less complete account of the history of the cosmos, from the supposed birth of space, time and matter out of the quantum vacuum until the emergence of life (at least on our planet Earth, and much probably elsewhere) and beyond.

# 1. From Myth to Myth

Questions about the origins of the universe, of the sky, of the earth, of life, of man have given rise to many different myths and legends and continue to be the subject of intensive research by astrophysicists, biologists and anthropologists (for details about the ancient cosmogonies referred to in the following, see e.g. Leeming, 2010). What were once fanciful stories are now scientific models but, whatever form they take, ideas about the origins of the universe both reflect and enrich the imagination of the people who generate them. Every society has developed its own stories to explain the creation of the world; most of them are ancient myths rooted in religion.

Whereas in monotheistic religions God is believed to have existed before the Creation, in most other kinds of religions the gods themselves are thought to originate from a creative element such as Desire, the Tree of the Universe, the Mundane Egg, Water, Chaos or the Void. Ideas like these appear in the *Rig-veda,* one of the four sacred books of the Brahmins and the oldest surviving written record of Indian culture which were compiled between 2000 and 1500 BC. The Tree of the Universe, symbol of the outward growth of the world and of its organic unity, is mentioned in ancient Indian legends as well as in those of the Babylonians and Scandinavians - who call it Yggdrasil. The anthropomorphic symbol of Desire was invoked by the Phoenicians and by the Maoris of New Zealand. The Mundane Egg, from which the Hindu Prajapatis (lords of all living things) emerged, also gave birth to the gods Ogo and Nommo, worshipped by the Dogon of Mali, and the Chinese giant Pan Gu as well as constituting the celestial vault in the legend of Orpheus.

A belief in some such primordial element, of which there are traces in every culture, underlies man's thinking about the history of the cosmos like a primitive universal symbol buried in the collective subconscious. This may explain the vague links which can always be discerned between this or that creation myth and modern scientific descriptions of the origin of the universe – for example, big bang theory. There is therefore nothing mysterious or surprising about these correspondences other than that certain ways of thinking about the world should be so ingrained in the human mind.

In fact scientific and mythical explanations of the Creation are neither complementary not contradictory; they have different purposes and are subject to different constraints. Mythical stories are a way of preserving collective memories that can be checked neither by the storyteller nor by the listener. Their function is not to explain what happened at the beginning of the world but to establish the basis of social or religious order, to impart a set of moral values. Myths can also be interpreted in many different ways. Science, on the other hand, aims to discover what really happened in historical terms by means of theories supported by observation. Often considered to be anti-myth, science has in fact created new stories about the origin of the universe: big bang theory, the theory of evolution, the ancestry of mankind. And many scientists, in their creative processes, implicitely invoke mythical images which may support (however tenuously) particular lines of thought. It is therefore hardly surprising that the new "creation" stories developed by scientists tend to be regarded by the general public as modern myths.

The theory of relativity, which is now the accepted framework within which both the structure and the evolution of the universe are described, is full of examples of this kind of mythologizing, despite the fact that it is 100 years old. Indeed the ideological basis on which Einstein built his 1917 model of a static eternal universe was partly philosophical (Einstein, 1917); in order to complete the structure, he had to invent a factor called the "cosmological constant" and incorporate it into his general theory of relativity. The discovery that the universe was expanding meant that Einstein's model had to be abandoned, and in 1931 Georges Lemaître proposed a scientific explanation of the birth of space and time (Lemaître, 1931; Luminet, 2011), according to which the universe resulted from the fragmentation of a "primeval atom" - a theory that recalls the ancient notion of everything hatching from a Mundane Egg. Lemaître's model was subsequently revised and adjusted and became the basis of modern big bang theory.

The concept of continuous creation, which enjoyed some popularity in the 1950s, is an even more striking example of scientific myth making. Big bang theory had not yet been fully substantiated by observation and many astrophysicists were reluctant to accept some of its metaphysical interpretations. Among them were Hermann Bondi and Thomas Gold who put forward the "steady state" theory (Bondi & Gold, 1948), whose fundamental principle, known as the "perfect cosmological principle", related back to Aristotle. The Greek philosopher had maintained that the world was eternal and indestructible and therefore without beginning or end (*De Caelo*, 279b 4-283b 22), in contradiction of Plato, who in *Timaeus* had expressed the idea that the world had begun and would end at a specific time. To compensate for the gradual dilution of matter that would result from the constant expansion of the universe, the steady state theorists had to invoke the idea of continuous creation of matter, at the approximate rate of one atom of hydrogen per cubic meter of space every five billion years. In the same year the English astronomer Fred Hoyle demonstrated that the steady state model was feasible on condition that a new field (which he called simply "C" for "creation") was added to the equation (Hoyle, 1948); this *ad hoc* invention was envisaged as a reservoir of negative energy which had existed throughout the life of the universe - i.e. for ever. The idea of continuous creation had appeared many times before (in the legends of the Aztecs, who believed that constant human sacrifices were necessary to regenerate the cosmos, as well as in the writings of Heraclitus and the Stoics, for example) and the scientific theory clearly followed this tradition. The steady state theorists, however, had to "force" their model to fit their philosophical views by introducing imaginary processes. The discovery of cosmic background radiation (Penzias & Wilson, 1965) finally disproved their hypothesis and provided evidence for Lemaître's big bang theory.

## 2. Chaos and Metamorphosis

The ancient Greeks had a great variety of myths relating to the history of the world. Although they all shared a language and a culture, each village, each tribe had its own beliefs, its own version of the Creation story and its own gods who were responsible for cosmic order.

Hesiod's *Theogony* (8th-7th century BC) was the first attempt to synthesize these traditions, which probably dated back to the Assyrian and Babylonian civilizations. In recounting the stages in the emergence of the gods from primordial chaos, *Theogony* offers an answer to the eternal questions of cosmogony: who created the world; what were the basic materials from which it was made; which came first, the gods, the stars or the elements?

Not only did *Theogony* have a strong influence on Greek thought, it also anticipated in some ways today's theories of the origin of the world - particularly the idea of primordial chaos. Since the universe appears to have an ordered structure (albeit an imperfect one), it seems logical to regard the state which preceded the Creation as one of disorder and confusion. This notion has provoked greater controversy than almost any other in the history of cosmogony.

Ovid's *Metamorphoses* also trawled Greek mythology, as well as Roman legend, in attempting to reconstruct the series of metamorphoses the world had undergone between the original state of Chaos and Julius Caesar's supposed transformation into a star: "Before the sea and the lands and the sky that covers all, / there was one face of nature in her whole orb / (they call it Chaos), a rough unordered mass, / nothing except inactive weight and heaped together /the

discordant seeds of unassembled things." (transl. Hill, 1985)

According to Hesiod the world was created *ex vacuo,* i.e. out of the void that existed before it, rather than *ex nihilo,* out of nothing. The distinction is fundamental. The standard Christian/Jewish interpretation of the Creation story in the Bible is that the world was created out of nothing, but the first few lines of *Genesis,* the *Book of Job* and the *Second Book of the Maccabees* give slightly different versions of the story - a discrepancy that has led to centuries of critical commentary. There are two possible interpretations of the scriptures: one that admits the pre-existence of some formless void (see e.g. St Augustine, *De Genesi Contra Manichaeos,* book I, chapter IV) and one that denies the existence of any matter before the Creation, the latter being the predominant belief in the Christian world today.

Modern historians have established that the *Genesis* story is a combination of two accounts written at different times. The later of these (*Genesis* 1 and 2, 1-4) was written after the flight from Israel and dates from the fifth or sixth century BC. Known as an elohistic text, it tells that the world was created in six days by the Elohim, the seventh day being a day of rest. The earlier account dates from between the ninth and the sixth century BC and is a jehovistic text in that it is the story of Jehovah, the God of Israel, who created man and woman to inhabit an earthly paradise (*Genesis* 2, 4-25). The elohistic story echoes the *Enuma Elish*, the Babylonian epic (18th century BC), in which the Creation begins with a struggle between God the Creator (whom the Babylonians called Marduk, the Jews Jehovah) and Chaos. Like the bible story it describes the Creation as a sequence of increasingly complex developments.

Quantum cosmology, which is now the most sophisticated scientific method of analyzing the history of the universe, is based on the general theory of relativity and quantum physics. It is an attempt to explain in mathematical terms how the universe suddenly emerged from the quantum vacuum (Brout et al., 1978; Vilenkin, 1984). Quite different from the traditional concept of the primeval void, the quantum vacuum is like a virtual ocean whose surface is continuously agitated by ripples of energy. These ripples can spontaneously generate pairs of particles and antiparticles, which disappear almost as soon as they appear, leaving behind a sort of bubbling brew of energy, in constant flux, called the "space-time foam". Occasionally the ripples create particles and antiparticles which are far enough apart not to cancel each other out. This is how matter emerges from the vacuum and how our entire universe could have arisen: from an ever expanding ocean of ripples.

## 3. Time and Creation

In any discussion of the creation of the world the paradoxical and complex question of temporality inevitably arises. If the Creation is regarded as an event, it must have taken place at some point in time, on a specific date. If time is regarded as a linear phenomenon, as it is in the Western world, this necessarily raises the problem whether anything existed before the Creation and, if so, what. But if time itself existed before the Creation, it cannot be part of the world as we know it - something which is difficult to imagine.

This paradox was pondered by Medieval scholars, who were forced to conclude that the world and time were created simultaneously. In the fourth century the Bishop of Milan, St Ambrose, wrote in his *Hexameron:* "In the beginning of time, therefore, God created heaven and earth. Time proceeds from this world, not before the world." (transl. Savage, 1961).

In the early 13th century the French philosopher and theologian William of Auvergne pursued a similar line of reasoning in his thinking about time: "Just as there is nothing beyond or outside the World, since it contains and includes all things, so there is nothing before or preceding time, which began with the creation of the World, since it contains all the periods of which it is comprised. This poses the question: What was before the beginning of time? or, since the word 'before' implies the existence of time, in the time preceding the beginning of time, did anything exist?" (transl. Teske, 1998).

The same questions continue to be asked today, and scientists who are asked to give public lectures on big bang theory and the expansion of the universe commonly face two kinds of questions: "What was there before the big bang?" and "What is there for the universe to expand into?" - in other words "Did time exist before time began?" and "Is there space beyond the limit of space?" The solution of modern physics to these paradoxes is that the universe consists of space-time and therefore the creation of the world cannot be regarded as a temporal phenomenon.

But before the development of quantum physics it was not possible to consider the origin of the universe from anything other than a temporal viewpoint. In the context of the Christian Creation story, for example, one of the fundamental questions was: did God create the world instantaneously or in stages? The various accounts of the Creation in the scriptures invite two different answers. The more widely accepted interpretation is that the Creation took place over six days (hexameron), followed by a day of rest. The theological aspects of pre-18th century Western cosmology derive in large part from the ancient hexamerons, commentaries on *Genesis* by the early Church Fathers. In the fourth century AD one of the greatest of these, St Basil of Caesarea, maintained that each day of the Creation corresponded to a normal day - i.e. the 24 hours between successive passages of the sun across the meridian - whereas, according to *Psalm* 90,4 "a thousand years in thy sight are but as yesterday".

The other interpretation of the bible story, that the Creation was instantaneous, was upheld in particular by Philo of Alexandria (*The Creation of the World*, chapter III), a contemporary of Jesus, and by Origen (*Contra Celsum*, chapters XXXVI-XXXVII.), like Philo a native of Alexandria, in the third century. How else could time have started before the appearance, on the fourth day, of the two great "astronomical clocks", the sun and the moon? Philo's argument was that, since a day is defined in terms of the passage of the sun across the sky, there could not have been any "days" of Creation before God made the sun. This concept of an instantaneous Creation can be seen as an attempt to escape if not a belief in temporality (this could hardly have been imagined at the time), at least its astronomical implications. In doing so these early philosophers anticipated certain aspects of modern cosmology, which, despite having more sophisticated means of recording time at its disposal, is still limited by them: since 1967 the duration of a second has been officially defined in terms of the radiation from caesium 133, but no such "natural clock" could have survived the extreme temperature and intense energy of the early universe; besides, caesium had not yet been "created"...

A few Christian philosophers attempted to reconcile these two apparently contradictory views of the Creation. St Thomas Aquinas (*Summa Theologiae*, question LXXIV), for example, argued that God had created the substance of things in an instant, but that he had taken six full days to accomplish the more complex task of separating, shaping and perfecting that substance. Once again this idea seems to anticipate modern thinking: according to big bang theory the history of the universe begins with the appearance of an entity called matter-space-time, which is followed by the separation of different particles (quark era, hadron era, radiation era, etc.). But, although modern cosmological theories can account for the gradual formation of matter, from that indeterminate "quark soup" to complex structures like galaxies, stars and planets, they are still unable to describe adequately the origin of the universe.

The distinction between origination and creation was made as long ago as the 13th century by Thomas Aquinas, who believed that the concept of creation could be explained rationally, whereas origination could be understood neither scientifically nor philosophically. "We may believe that there was a beginning to the World, but we can neither prove it nor explain it," he wrote in his *Summa Theologiae* of 1266-73. In another work *(De Aeternitate Mundi)* he even refuted the argument that the world had begun at a particular time and defended the possibility that the universe, although it had been created, was eternal. Other medieval thinkers took up this idea. Albertus Magnus (c. 1200-80), for example, maintained in his *Physica* that the beginning of the World was not a physical act of creation and cannot be proven as such.

Such questions continue to exercise the minds of cosmologists everywhere, albeit within the very different conceptual framework of big bang theory. This theory is often wrongly believed to describe the origin of the universe. In fact, although it does account for the evolution of the universe from soon after its birth (estimated to have been 13.8 billion years ago), it does not claim to be able to envisage a "zero time", nor a fortiori the moment of creation or the origin of the universe. The extreme conditions out of which space, time, light and matter must have arisen remain beyond the reach of scientific investigation, since at the theoretical "zero time" the temperature would have been at infinity and modern physics is incapable of describing any interaction occurring at more than $10^{32}$ Kelvin (known as Planck temperature). The energy generated at higher temperatures is such that quantum laws would take place at the very core of the structure of space-time, making it impossible to calculate the physical effects of that activity within any current theory. There is thus a "barrier" preventing us from uncovering the history of the universe beyond a certain point; our understanding cannot reach "zero time" but is limited to the end of Planck time, $10^{-43}$ seconds later. In fact those first instants in the life of the universe are best described in terms of energy and temperature rather than time.

## 4. The Date of the Creation

None of the traditional myths gives a precise date for the Creation. The very idea of putting dates to the history of the world seems to have been foreign to the mentality of the ancients. For them the origin of the universe was simply a notion which helped to understand the separation of reality into two regions: formless chaos and cosmic order. It was the Jewish/Christian preoccupation with time as a linear process which prompted the question: when was the Creation? From then on the greatest theologians (from Eusebius of Caesarea in the fourth century to James Ussher, Irish prelate and archbishop of Armagh, in the 17th century) and scientists (from Kepler to Newton) would attempt to provide the answer.

For centuries the only clues were to be found in the bible, which was thought to be able at least to provide an upper limit to the age of the world. Thus the vast majority of scholars put the date of the Creation at around 4000 BC, the most common method of calculation being to count the number of generations between Adam and Jesus.

Funnily enough, more precise estimates gradually appeared. In his historical treatise *Annales Veteris Testamenti, a Prima Mundi Origine Deducti (Annals of the Old Testament, Traced Back to the Origin of the World)* of 1650, James Ussher attempted to determine precisely the dates of the great biblical events by checking them against historical facts and astronomical phenomena. According to his calculations the first day of the Creation was 23rd October 4004 BC (beginning at midday).

These speculations were even incorporated into a new edition of the bible which appeared in 1701, but doubt was soon cast on them by a number of archaeological discoveries: it was found that there were already established civilisations in the great cities of Egypt in 4000 BC and similarly advanced societies existed in parts of Asia... The date of the Creation, which had to predate the appearance of the human race, would have to be put back.

The hundred years after the death of Newton saw the birth of mechanism - a philosophical theory which explained all natural phenomena in terms of combinations of physical actions - and scientific rationalism - the belief in natural wisdom as opposed to wisdom resulting from divine revelation. The reconstruction of the past by purely rational methods was as yet too shaky a foundation on which to base a Creation date other than that suggested by the traditional interpretation of *Genesis*. But the development of geology and the discovery of fossils in the Alps in the early 18$^{th}$ century made more and more difficult to sustain the biblic datation.

The first scholars seriously to question the diluvian theory were the French archaeologist Benoit de Maillet and his compatriot, the naturalist Georges Louis Leclerc, comte de Buffon. In his *Telliamed,* which was originally published secretly, Maillet anticipated Darwin by 100 years in proposing that the world had evolved rather than been created in its final form and particularly in imagining that life had originated in the sea and that the earth was several million years old. Buffon, in his *Théorie de la Terre (Theory of the Earth)* of 1749, used the results of physical experiments as evidence of the age of the earth: he heated cannonballs until they were red hot, then measured the time they took to cool and concluded that the earth must be 74,832 years old.

During the 19th century the age of the earth, and of the solar system in general, was progressively increased as a result of discoveries in the new science of thermodynamics as well as in astronomy, geology and paleontology. Previously the human race had been assumed to be as old as the earth, which had been created for man to live in. From now on it was recognised that, in relation to the age of the earth, the whole of human history was like the blink of an eye in an hour, and it became increasingly difficult to believe that humanity was the *raison d'être* of the earth and the crowning achievement of God's Creation.

Today big bang theory tells us the "age" of the universe - more exactly the duration of cosmic evolution - which is the length of time since the end of the Planck era, 13.8 billion years. This is calculated by measuring the rate of expansion of the universe, in other words the speed at which galaxies are moving away from each other.

Now the exact age of the universe is known, the chronology of its evolution (decoupling of fundamental forces, appearance of elementary particles, star formation, etc.) is also firmly established, in some cases with mind-boggling accuracy. At the incredibly high temperatures immediately after the big bang things happened very quickly and almost as many changes took place in the first millionth of a second after the Planck era as took place in the next billion years. A few million years here or there in the life of a star is therefore less significant than a few billionths of a second in the life of an elementary particle...

## 5. Creation vs. Evolution

The ancient Babylonians had a different idea of how the world began. They believed that it had evolved rather than being created instantaneously. Assyrian inscriptions have been found which suggest that the cosmos evolved after the Great Flood and that the animal kingdom originated from earth and water. This idea was at least partially incorporated into a monotheist doctrine and found its way into the sacred texts of the Jews, neighbours and disciples of the Babylonians. It was also taken up by the early Ionian philosophers, including Anaximander and Anaximenes, and by the Stoics and atomists. Democritus developed a theory that the world had originated from the void, a vast region in which atoms were swirling in a whirlpool or vortex. The heaviest matter was sucked into the centre of the vortex and condensed to form the earth. The lightest matter was thrown to the outside where it revolved so rapidly that it eventually ignited to form the stars and planets. These celestial bodies, as well as the earth itself, were kept in position by centrifugal force. This concept admitted the possibility that the universe contained an infinite number of objects. It also anticipated the 19th century theory of the origin of the solar system, known as the nebular hypothesis, according to which a "primitive nebula" condensed to form the sun and planets.

The idea of universal evolution had a strong influence on classical thought and developed in various directions during Greek and Roman times. In the first century BC Lucretius (*De Natura Rerum,* book V) extended the theories of atomism and evolution to cover every natural phenomenon and argued that all living things originated from earth . The Church Fathers, who insisted that the Creation was instantaneous, rejected any sort of evolutionary theory; to them the ideas of the Stoics and atomists were heretical.

In the second half of the 16th century the idea of universal evolution began to be incorporated into the new system of scientific thought resulting from the work of Copernicus, Kepler, Galileo, Descartes and Newton. According to Descartes, for example, space consisted of "whirlpools" of matter whose motion was governed by the laws of physics. Newton, with his theory of universal attraction, was accused of having substituted gravitation for providence, for having replaced God's spiritual influence on the cosmos by a material mechanism (on this controversy see for example McCosh, 1890). A new view of the world had nevertheless been established, whereby the workings of the universe were subject not to the whim of the Almighty but to the laws of geometry and physics - it was an irreversible step.

Indeed, the idea that the visible world is the manifestation of an underlying mathematical order goes back to Pythagoras school of thought. Plato's *Timaeus* already described the creation of the cosmos as a process of bringing order and harmony to what was previously formless, and promoted the idea that the act of Creation must have been guided by some overriding geometric principles. This was taken up by the Neo-Platonist philosophers of the Renaissance and influenced many 17[th] century thinkers. The "geometric Creator" often appears in medieval iconography with a pair of compasses in his hand, as described by Milton in his epic poem *Paradise Lost* of 1667, and later in a famous illuminated printing by William Blake in 1794. As science developed, the concept of a mathematically determined Creation became more firmly established. Thomas Wright (Wright, 1750) still located a divine creative influence at the centre of each galaxy, but most 18th century cosmogonies played down the role of the Creator. When asked by Napoleon what God's role was in the Creation, the French mathematician and astronomer Pierre Simon Laplace, author of the monumental *Mécanique céleste* (1798-1825), replied:

"Sir, I have not needed that hypothesis." (quoted e.g. in Luminet, 2008).

The Pythagorean/Platonic belief that the creation, or at least the construction, of the universe can be described in mathematical terms is still current today, even if we now refer to "group theory" instead of numbers or geometric shapes.

## 6. Conclusion : a Modern Cosmogenesis

In the first quarter of the 20th century cosmology became a distinct scientific discipline, thanks in part to the theoretical advance made in 1915 by Einstein with his general theory of relativity and in part to the revolution in observational techniques which revealed the true extent of the universe (for an historical account of the development of relativistic cosmology, see e.g. Luminet, 2006).

Big bang theory is now the established historical framework for the study of the universe and today's astrophysicists claim to be able to give a plausible account of its 13.8 billion years history right back to a micro-second after its supposed birth. The universe has undergone a gradual process of expansion and cooling ever since the big bang; at the same time increasingly complex physical structures have evolved.

The past history of the universe can conveniently be divided into two main periods: the first million years - infancy - and the remaining 13.8 billion years - maturity - (for a more detailed description, still pedestrian, see e.g. Glendenning, 2004).

During the Planck era, time and the dimensions of space as we know them were so intimately linked as to be practically indistinguishable. Various speculative theories of quantum cosmogenesis, as yet in their infancy, attempt to explain how our universe emerged at the end of the Planck era. Some physicists refer to its "spontaneous emergence", others to an infinite number of separate "cosmic bubbles" arising from the quantum vacuum like foam from the surface of the sea.

Between $10^{-43}$ and $10^{-32}$ seconds after the big bang the infant universe consisted of elementary particles bound by a primeval superforce. A few billiseconds later gravity separated itself from the surviving electrostrong force, which in turn, as the temperature fell to $10^{27}$ degrees, divided into the strong force and the electroweak force. Experiments in high energy physics suggest that these "symmetry breakdowns" had spectacular consequences, such as the appearance of strange fields and particles (e.g. the Higgs-Englert boson), or the onset of "inflation" - a very short period during which the universe grew immeasurably. The fundamental constituents of matter - quarks, electrons and neutrinos - also appeared at this time.

$10^{-11}$ seconds after the big bang the temperature of the universe had dropped to $10^{15}$ degrees and the electroweak force split into an electromagnetic and a weak force, thus establishing the four fundamental forces and fixing the physical conditions for the formation of complex structures.

$10^{-6}$ seconds after the big bang all quarks were "linked" in threes by the strong force to form the first nucleons, i.e. protons and neutrons. By this time the temperature had fallen to a billion degrees as the universe continued to expand. As particles became more widely spaced, they collided less frequently but one hundred seconds or so later the crucial process of nucleosynthesis began. Neutrons and protons combined to form the simplest atomic nuclei: hydrogen, helium and lithium (in various isotopes). Most of the universe, however, remained as isolated protons, i.e. as hydrogen nuclei.

Nucleosynthesis took place only for a very short time: the universe was cooling so rapidly that there was only time for the lightest elements to form. These therefore constitute 99 per cent of the visible matter in the universe today (75% hydrogen and 24% helium in mass fraction). The remaining one per cent, consisting of heavier elements like carbon, nitrogen and oxygen, would not be created until billions of years later, when the stars were formed.

Until it was 380,000 years old the universe remained opaque; in other words it emitted no radiation: the density of electrons prevented photons from moving freely. But the universe, consisting of a "soup" of particles and radiation, continued to cool and expand until, at 3,000 degrees, it became transparent and emitted its first electromagnetic signal in the form of what we now detect as cosmic background radiation.

A million years after the big bang the first atoms were formed, when electrons were captured by hydrogen and helium nuclei, and these atoms combined into molecules to create vast clouds of hydrogen, out of which stars would later emerge.

A billion years after, the first galaxies were formed. One of these was probably our own Milky Way, in which several generations of stars have since come and gone. Cosmic gases condensed to form the sun about nine billion years later, i.e. about five billion years ago. Within a relatively short time the planets solidified around it, the most reliable figure for the age of the earth being 4.56 billion years. Once the initial intense meteorite bombardment had ended and the earth had cooled, life began to appear in the oceans: single cell organisms first developed 3.5 billion years ago (may be earlier). From then on the pace of evolution accelerated: the first vertebrae appeared 600 million years ago, the first mammals 200 million years ago. Our own species, Homo Sapiens, developed only recently - two million years ago.

Such a story is not a statement of atheism, rather a simple recognition that the question "What about God ?" lies outside of the field of science. Modern cosmology contents itself with reconstructing the present and past events of cosmic history, starting from observations, laboratory experiments and the theoretical models believed to best represent the Universe. Within the framework of big bang models, it tries, as closely as possible, to approach the conditions that might have presided at the "appearance" of space, time, and matter during an event extrapolated into the past, the big bang. The mathematical nature of the big bang - that of a singularity where the curvature of space-time is infinite - implies in fact that the big bang does not belong to the space-time geometry. The big bang is therefore not even an event. It has not taken place and has no location. Similarly, it necessarily lies outside the domain of our present-day theories.

In addition, new "scenarios" from quantum gravity theories smooth out the big bang as a singularity and tentavively describe a pre-big bang era for the universe. Thus it is fallacious as well as naive to see in the mysterious and inaccessible big bang a metaphor of Creation or the "Mind of God" (Davis, 1992). Physics does not serve to reveal the attributes and intentions of an alleged Creator, rather it provides a means for better understanding nature. The authentic cosmological question of knowing if the Universe or matter have a temporal origin has often been transformed into a problem of creation. This shift rests on the mistaken idea that creation necessarily requires an exterior agent, a cause external to the physical world. This confusion serves as a basis for the reactions of adversaries of the big bang model, for the metaphysical drifting of its partisans, and for attempts at recovery by theologians.